\documentclass[prl,twocolumn,amsmath,amssymb,nobibnotes,nofootinbib,preprintnumbers]{revtex4}

\usepackage{graphicx}% Include figure files
\usepackage{bm}% bold math
\usepackage{xcolor}

\begin{document}

\preprint{YITP-13-67}

\title{Hartle-Hawking no-boundary proposal in dRGT massive gravity:\\
Making inflation exponentially more probable}

\author{Misao Sasaki}
 \email{misao_AT_yukawa.kyoto-u.ac.jp}
\author{Dong-han Yeom}%
 \email{innocent.yeom_AT_gmail.com}
\author{Ying-li Zhang}%
 \email{yingli_AT_yukawa.kyoto-u.ac.jp}

\affiliation{Yukawa Institute for Theoretical Physics, Kyoto University,
Kyoto 606-8502, Japan}%

\date{\today}% It is always \today, today,

\begin{abstract}
It is known that the no-boundary proposal in the traditional
Einstein gravity does not prefer inflation, that is, the probability
of realizing a large number of $e$-folds is exponentially
suppressed. This situation may be changed drastically in a class of
nonlinear massive gravity theories recently proposed by deRham,
Gabadadze and Tolley, called dRGT massive gravity. We show that the
contribution from the massive gravity sector can enhance the
probability of a large number of $e$-folds substantially for a
sufficiently large mass parameter $m_g$ comparable to the Hubble
parameter during inflation, say $m_g\gtrsim 10^{12}$GeV. We
illustrate possible models to trigger such a large mass parameter in
the early universe while it is negligibly small in the present
universe. This opens a new window to explore the inflationary
scenario in the context of quantum cosmology.
\end{abstract}

\pacs{Valid PACS appear here}% PACS, the Physics and Astronomy
\maketitle

\section{Introduction}

 It is known that quantum gravity is hoped to
resolve the initial singularity of the universe and assign a proper
initial condition with a reasonable probability, where the
corresponding wavefunction of the universe is discribed by the
Wheeler-DeWitt equation \cite{DeWitt:1967yk}.
%Although it is not
%well-defined in the rigorous sense due to the lack of full quantum
%theory of gravity, one expects that it is valid at least at the
%semi-classical level. Then the leading order solution may be
%constructed in the mini-superspace in which the only dynamical
%variable of the metric is the cosmic scale factor of a homogeneous
%and isotropic universe.
In order to obtain a unique solution to this equation, one natural
idea to fix the boundary condition is to choose the initial state
such that it most closely resembles the ground state, or the
Euclidean vacuum state. Moreover, it is known that the Euclidean
vacuum state wavefunction in standard quantum theory can be obtained
by the Euclidean path integral. Hence, Hartle and Hawking
\cite{Hartle:1983ai} generalized it to the case of gravity, and
introduced the Euclidean path integral to choose the wavefunction of
the universe,
\begin{eqnarray}
\Psi[h_{ij}, \chi] = \int_{\partial g = h, \partial \phi = \chi}
\mathcal{D}g\mathcal{D}\phi \;\;e^{-S_{\mathrm{E}}[g,\phi]},
\label{Psi}
\end{eqnarray}
where $h_{ij}$ is the metric of a compact 3-geometry $\Sigma$
and $\chi$ is the matter (inflaton) field $\phi$ on $\Sigma$,
which form the boundary of all possible, regular
 4-geometries and matter configurations over which the path integral
is to be performed. This is called the \textit{no-boundary proposal}.
However it is known that the Hartle-Hawking no-boundary
proposal is in severe conflict with the realization of successful inflation \cite{Vilenkin:1987kf}.

It has been proved that an $O(4)$-symmetric solution gives the
lowest action for a wide class of sc alar-field theories, hence
dominating the path integral (\ref{Psi}). So it is reasonable to
assume that the metric owns the $O(4)$ symmetry even in the presence
of gravity \cite{Coleman:1977th},
\begin{eqnarray}
ds_{\mathrm{E}}^{2} = d\tau^{2} + a^{2}\left(\tau\right) d\Omega_{3}^{2}.
\end{eqnarray}
On the other hand, under the steepest-descent approximation (or
equivalently, WKB approximation) where the wavefunction is dominated
by sum-over on-shell histories, the corresponding actions are
complex-valued,
\begin{equation}
\Psi[\tilde{a},\tilde{\phi}]
\simeq A[\tilde{a},\tilde{\phi}] e^{i S[\tilde{a},\tilde{\phi}]},
\label{eq:class}
\end{equation}
where $\tilde{a}$ and $\tilde{\phi}$ are the boundary values of $a$
and $\phi$, respectively, and $A$ and $S$ are real-valued functions.
When the phase $S$ varies much faster than $A$,
\begin{equation} \label{eqn:classicality}
\left|\nabla_I A[\tilde{a},\tilde{\phi}]\right|\ll
\left|\nabla_I S[\tilde{a},\tilde{\phi}]\right|, \qquad I=\tilde{a},\tilde{\phi},
\end{equation}
these histories are classical since it satisfies the semi-classical
Halilton-Jacobi equation \cite{Hartle:2008ng}.

Thus, provided that the potential is exactly flat, if the values
$\tilde{a}$ and $\tilde{\phi}$ are such that they form a maximal
slice of an $O(4)$ symmetric regular Euclidean solution, that is,
$\tilde{a}$ and $\tilde{\phi}$ are the values at which
$\dot{{a}}=\dot{\phi}=0$, then the probability of the existence of
such a classical universe is evaluated as
\begin{equation}
P \simeq e^{- 2 S_{\mathrm{E}}},
\label{probability}
\end{equation}
where $S_{\mathrm{E}}$ is half the Euclidean action of the solution
with the maximal slice at which $a=\tilde{a}$ and $\phi=\tilde{\phi}$.
At the maximal slice the Euclidean solution can be analytically continued
to a Lorentzian classical solution, and a classical universe is born
with its initial data given by $a=\tilde{a}$ and $\phi=\tilde{\phi}$.

This analysis can be applied to evaluate the probability of creation
of a universe that has witnessed an early inflationary era, at which
the potential satisfies the slow-roll conditions. For simplicity,
let us consider the chaotic inflationary scenario where the
potential $V(\phi)\propto\phi^n$ with $n>0$ has reflection symmetric
around $\phi=0$ and monotonically increases as $|\phi|$ increases.
In Einstein gravity, there is a continuous distribution of
complex-valued instantons for $|\phi|> \phi_{\mathrm{cut}}\sim 1$
(hereafter, we adopt the natural units $8\pi
G=M_{\mathrm{Pl}}^{-2}=1$), where $\phi_{\mathrm{cut}}$ is a cutoff
scale below which the classicality condition is no longer
satisfied~\cite{Hartle:2008ng,Hartle:2007gi,Hwang:2011mp}. When
there are approximately classical Euclidean solutions, the slow-roll
condition is generally satisfied in the Lorentzian regime, and the
Euclidean action is approximately given by
\begin{equation}\label{SEGR}
S_{\mathrm{E}} \simeq -\frac{12\pi^{2}}{V(\phi)}\,.
\end{equation}
Inserting this into Eq.~(\ref{probability}), one immediately obtains
the probability of creation of such a universe:
\begin{equation}\label{PN}
P\propto\exp\left(\frac{24\pi^2}{V(\phi)}\right)\quad\Longrightarrow\quad
\ln P\propto N^{-\frac{n}{2}}\,,
\end{equation}
where $N$ is the corresponding e-folding number defined by $dN=Hdt$.
It is obvious from Eq.~(\ref{PN}) that an early universe with a
large vacuum energy or large number of $e$-folds is exponentially
suppressed.
Hence, a successful inflationary scenario
with sufficient e-folding number (e.g. 50-60 e-folding numbers) is
disfavored, which is thought to be a defect of the no-boundary
proposal
 (for alternative explanations and reviews, see
 \cite{Hwang:2012bd}). However, this severe problem can be naturally solved in a modified gravity
theory with a non-vanishing graviton mass, that is, the so-called
massive gravity theory.

\section{dRGT Massive gravity}

Recently, with various motivations, massive gravity models have been
extensively studied. In this paper, we consider nonlinear massive gravity
proposed by deRham, Gabadadze and Tolley~\cite{deRham:2010kj}, the
so-called dRGT massive gravity,
\begin{eqnarray}
S = \int \sqrt{-g}~d^4x \left[\frac{R}{2} +
\mathcal{L}_{\mathrm{mg}} - \frac{1}{2} \left(\nabla \phi\right)^{2}
- V(\phi) \right],
\end{eqnarray}
where $\phi$ is the inflaton and $\mathcal{L}_{\mathrm{mg}}$ is the massive gravity term
given by
\begin{eqnarray}
\mathcal{L}_{\mathrm{mg}}
 &=& m_g^2\Biggl[\frac{1}{2} \left( \left[ \mathcal{K} \right]^{2}
- \left[ \mathcal{K}^{2}\right] \right) \nonumber \\
&+& \alpha_{3}\frac{1}{6} \left( \left[ \mathcal{K} \right]^{3}
- 3\left[ \mathcal{K} \right]\left[ \mathcal{K}^{2}\right]
+ \left[ \mathcal{K}^{3}\right] \right) \nonumber \\
&+& \alpha_{4}\frac{1}{24} \Bigl( \left[ \mathcal{K} \right]^{4}
 - 6 \left[ \mathcal{K} \right]^{2} \left[ \mathcal{K}^{2}\right]
+ 3 \left[ \mathcal{K}^{2}\right]^{2} \nonumber \\
&& \qquad+ 8  \left[ \mathcal{K} \right]\left[ \mathcal{K}^{3}\right]
- 6\left[ \mathcal{K}^{4}\right] \Bigr)\Biggr]\,,
\end{eqnarray}
with $m_g^2$ being a graviton mass parameter, $\alpha_3$
and $\alpha_4$ being non-dimensional parameters, and
\begin{eqnarray}
\mathcal{K}^{\mu}_{\nu} = \delta^{\mu}_{\nu}
- \sqrt{g^{\mu\sigma}G_{ab}(\varphi)\partial_{\nu}\varphi^{a}
   \partial_{\sigma}\varphi^{b}}.
\end{eqnarray}
The fields $\varphi^a$ ($a=0,1,2,3$) are called the St\"{u}ckelberg
fields whose role is to recover the general covariance.

We set the fiducial, field space metric $G_{ab}(\varphi)$ to be
a de Sitter metric~\cite{Zhang:2012ap,dS:2012},
\begin{eqnarray}
G_{ab}(\varphi)d\varphi^{a}d\varphi^{b} \equiv
- (d\varphi^{0})^{2} + b^{2}(\varphi^{0})d\Omega_{3}^{2},
\end{eqnarray}
where $b(\varphi^0) = F^{-1} \cosh (F \varphi^0)$ with $F^{-1}$
being the curvature radius of the fiducial metric. For simplicity,
we assume $G_{ab}(\varphi)$ to be non-dynamical as in the original
dRGT gravity~\cite{deRham:2010kj}. However, the theory may be more
consistently formulated if $G_{ab}$ is made
dynamical~\cite{Hassan:2011zd}. We will not discuss
further this issue here since it is beyond the scope of this short
article.

Under the assumption of $O(4)$ symmetry, we set $\varphi^0=f(\tau)$.
Then after a straightforward calculation, we obtain $b(\varphi^0)=
X_{\pm} a(\tau)$, where~\cite{Gumrukcuoglu:2011ew}
\begin{eqnarray}
X_{\pm} \equiv
\frac{1 + 2\alpha_{3} + \alpha_{4} \pm \sqrt{1 + \alpha_{3}
+ \alpha_{3}^{2} - \alpha_{4}} }{\alpha_{3} + \alpha_{4}}\,.
\end{eqnarray}
We note that $X_{\pm}$ must be positive. This constrains the
parameter space of the theory.
The equations of motion are given by
\begin{eqnarray}
\dot{a}^{2} - 1 - \frac{a^{2}}{3}
\left( \frac{\dot{\phi}^{2}}{2} - V_{\mathrm{eff}} \right) &=& 0\,,\\
\ddot{\phi} + 3 \frac{\dot{a}}{a} \dot{\phi} - V'_{\mathrm{eff}} &=& 0\,,
\end{eqnarray}
where a dot ($\dot{~}$) denotes $d/d\tau$, a prime (${~}'$) denotes
$d/d\phi$, and the effective potential is given by
\begin{eqnarray}
V_{\mathrm{eff}} (\phi) = V(\phi) + \Lambda_{\pm}\,,
\end{eqnarray}
with $\Lambda_{\pm}$ being the cosmological constant due to
the massive gravity terms,
\begin{eqnarray}
&&\Lambda_{\pm}
=- m_g^{2} \left( 1 - X_{\pm} \right) \Bigl[ 3\left(2 - X_{\pm} \right)
 \nonumber \\
&&\quad
+ \alpha_{3} \left(1 - X_{\pm} \right)\left(4 - X_{\pm} \right)
+ \alpha_{4} \left(1 - X_{\pm} \right)^{2} \Bigr]\,.
\end{eqnarray}
Using these equations of motion, we obtain the on-shell action
as~\cite{Zhang:2012ap}
\begin{eqnarray}
S_{\mathrm{E}} = 2 \pi^{2} \int d\tau \left[ 2a^{3} V_{\mathrm{eff}}
 - 6 a - m_g^{2} a^{3} Y_{\pm} \sqrt{-\dot{f}^{2}} \right]\,,
\end{eqnarray}
where
\begin{eqnarray}
Y_{\pm} \equiv  3 (1-X_{\pm})
+ 3 \alpha_{3} (1-X_{\pm})^{2} + \alpha_{4} (1-X_{\pm})^{3}.
\end{eqnarray}

Now we assume that the potential is sufficiently flat so that
at leading order approximation we can set $V_{\mathrm{eff}}'=0$,
hence $\dot\phi=0$. Thus we obtain
\begin{eqnarray}
\phi &=& \phi_{0}\,,
\\
a &=& \frac{1}{H} \cos H\tau,
\end{eqnarray}
where $H^{2} = V_{\mathrm{eff}}(\phi_0)/3$.
Then the action (over the half hemisphere) is given by~\cite{Zhang:2012ap}
\begin{eqnarray}\label{SEMG}
S_{\mathrm{E}} &\simeq&
%-\frac{4\pi^{2}}{H^{2}}
%\left( 1 - \frac{m_g^{2}}{H^{2}} X_{\pm}Y_{\pm} C\left(\alpha^2 \right) \right)
%\cr
%&=&
-\frac{4\pi^{2}}{H^{2}}
\left( 1 - \frac{m_g^{2}}{F^{2}} \frac{Y_{\pm}}{X_{\pm}}
\alpha^2C\left(\alpha^2\right) \right)\,,
\label{SE}
\end{eqnarray}
where $\alpha = X_{\pm}F/H$ and
\begin{eqnarray}\label{C}
C\left(\alpha^2 \right) \equiv
\frac{2 - \sqrt{1 - \alpha^{2}} \left(2 + \alpha^{2} \right)}{6 \alpha^{4}}\,.
\end{eqnarray}
Comparing Eq.~(\ref{SEMG}) to~(\ref{SEGR}), using the Friedman
equation $3H^2=V(\phi)$, one finds that a counter term proportional
to $m_g^2$ appears in dRGT massive gravity theory, which drastically
changes the behavior of wavefunction so that the distribution of
probability may not peak at $H^2\simeq0$. It should be noted that
the above approximation is valid as long as the slow-roll condition
is satisfied and the field value is well outside the cutoff, i.e.,
$|\phi_0|> \phi_{\mathrm{cut}}$.

\begin{figure}
\begin{center}
\includegraphics[scale=0.7]{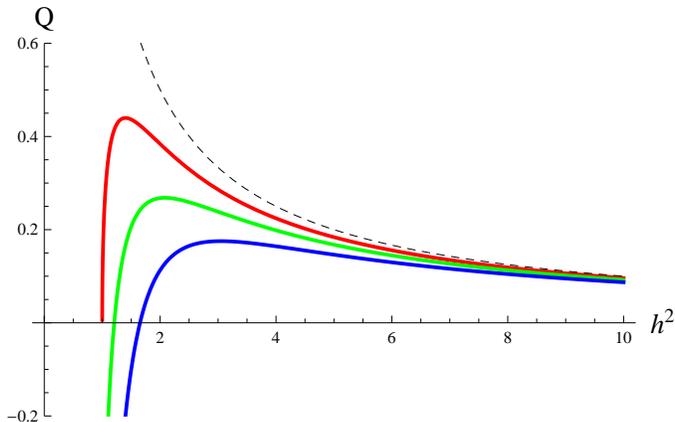}
\caption{\label{fig:plotQ}The function $Q(u)$ as a function of the
normalized Hubble parameter $h^2=1/u=H^2/(X_{\pm}^2F^2)$ for
$m_g^2Y_{\pm}/(F^2X_{\pm})=0$ (dashed), $3$ (red), $6$ (green), and
$10$ (blue) from top to bottom. As has been expected, in traditional
Einstein gravity where $m_g=0$, the probability will exponentially
decrease when $H^2$ increases, which implies disfavor of inflation
with a large number of e-foldings. However, in dRGT massive gravity,
counter terms proportional to $m_g^2$ appear so that the probability
peaks at much larger $H^2$. This implies the preference of a much
larger e-folding number for a successful inflationary scenario,
hence offers a way to realize inflation in the context of quantum
gravity.}
\end{center}
\end{figure}
%\begin{figure}
%\begin{center}
%\includegraphics[scale=0.65]{plotG1}
%\caption{\label{fig:plotG}The function $G(u)$ as a function of
%$u=\alpha^2$. The thin gray line is the line
%$G_{\mathrm{approx}}=u/2$ for reference.}
%\end{center}
%\end{figure}

\section{Sufficient $\bm{e}$-folding number for inflation}
As mentioned above, we must have $X_{\pm}>0$. On the other hand,
$\alpha$ must be in the range $0<\alpha<1$~\cite{Zhang:2012ap}.
Moreover, for definiteness, we also assume $Y_{\pm}>0$. From
Eq.~(\ref{C}), $C(\alpha)>0$, which implies that the absolute value
of $S_E$ (or $-S_E$) is smaller in the massive gravity case than in
the Einstein case, provided with the same value of $H$.

Alternatively, if we fix the model parameters and vary $H$,
$\alpha^2C(\alpha^2)\to0$ as $H\to\infty$ while it approaches 1/3
as $H\to H_{\mathrm{min}}$ where $\alpha(H_{\mathrm{min}})=1$
or $H_{\mathrm{min}}=X_{\pm}F$.
Then there arises a hope that the probability of a universe with
small $H$ may be substantially suppressed, or conversely the
probability of a universe with larger $H$ is exponentially enhanced.

To see if this is the case or not, let us introduce variable
$u\equiv\alpha^2=X_{\pm}^2F^2/H^2$ and rewrite Eq.~(\ref{SEMG}) in
the following form:
\begin{eqnarray}
S_{\rm
E}&=&-\frac{4\pi^2}{X_\pm^2F^2}\left[u-\frac{m_g^2}{F^2}\frac{Y_{\pm}}{X_{\pm}}u^2C(u)\right]\cr
&\equiv&-\frac{4\pi^2}{X_\pm^2F^2}Q(u)\,,
\end{eqnarray}
where $Q(u)\equiv -X_\pm^2F^2S_{\rm E}/(4\pi^2)$. Then one finds
\begin{eqnarray}
\frac{\partial Q}{\partial H^2}
&=&-\frac{\alpha^2}{H^2}\frac{\partial Q}{\partial\alpha^2}
\cr
&=&-\frac{\alpha^2}{H^2}\left(
1-\frac{m_g^2}{F^2}\frac{Y_{\pm}}{X_{\pm}}\frac{\alpha^2}{4\sqrt{1-\alpha^2}}
\right)\,.
\end{eqnarray}
The function $Q(u)$ is plotted in Fig.~\ref{fig:plotQ} as a function
of the normalized Hubble parameter $h^2=1/u$, for
$m_g^2Y_{\pm}/(F^2X_{\pm})=0$, $3$, $6$, and $10$, respectively. It
is readily seen that unlike the case for traditional Einstein
gravity where $m_g=0$ (dashed line), in dRGT massive gravity theory,
the function $Q$, hence $-2S_E$ will be maximized at
$\alpha^2=\alpha_{\mathrm{m}}^2$ where
\begin{eqnarray}
\frac{\alpha_{\mathrm{m}}^2}{\sqrt{1-\alpha_{\mathrm{m}}^2}}
=\frac{4F^2}{m_g^2}\frac{X_{\pm}}{Y_{\pm}}\,.
\end{eqnarray}
Since the left-hand side varies monotonicallly
from zero to infinity as $\alpha^2$ varies in the range $0<\alpha^2<1$,
there always exists a unique maximum provided that both $X_{\pm}$
and $Y_{\pm}$ are positive.
Thus the probability is maximized at $\alpha^2=\alpha_{\mathrm{m}}^2$
with the exponent given by
\begin{eqnarray}
\ln
P&\approx&-2S_{\mathrm{E}}=\frac{8\pi^2}{X_{\pm}^2F^2}G(\alpha_{\mathrm{m}}^2)\,;
\cr \cr G(u)&\equiv& \frac{u^2-2u+4-4\sqrt{1-u}}{3u}\,, \label{lnP}
\end{eqnarray}
which may be well approximated by $G(u)\approx u/2$ when $u\ll1$.

We see that the problem of the no-boundary proposal may be solved
for a sufficiently wide range of the parameter space. For example,
assuming both $X_{\pm}$ and $Y_{\pm}$ are of order unity, we may
consider the case $m_g^2\gg F^2$ which implies
$\alpha_{\mathrm{m}}^2\approx4 X_{\pm}F^2/(Y_{\pm}m_g^2)\ll1$.
Inserting this into Eq.~(\ref{lnP}), one finds
\begin{eqnarray}
\ln P\approx \frac{16\pi^2}{m_g^{2}X_{\pm}Y_{\pm}}
\approx\frac{4\pi^2}{H_{\mathrm{m}}^2}\,,
\end{eqnarray}
where $H_{\mathrm{m}}^2\approx X_{\pm}Y_{\pm}m_g^2/4$
is the Hubble parameter at which the probability is
maximized. Thus a classical universe emerges most probably
with the Hubble parameter $H^2=O(m_g^2)$. For $m_g^2$
close to the Planck scale this gives sufficient inflation.

\section{Triggering mass parameter}

We have shown that for a sufficiently large mass parameter $m_g$
the problem associated with the no-boundary proposal that
the expected number of $e$-folds is too small to realize
successful inflation may be solved. However, since we know that the
graviton mass should be extremely small today, we need a mechanism
to make it large only in the very early universe.

Here we present a couple of speculations for such a mechanism.
\\
\noindent (1) \textit{Field-dependent mass}: Let us consider the
case when $m_g^2$ is a function of the inflaton
$\phi$~\cite{Huang:2012}.
\begin{eqnarray}
m_g^2=m_g^2(\phi)\,.
\end{eqnarray}
If $m_g^2$ is finite for $|\phi|>\phi_{\mathrm{cut}}$ but
exponentially small
 for $|\phi|<\phi_{\mathrm{cut}}$, then our analysis in the case of massive
gravity is still valid and the fact that there is no classical
histories at $|\phi|<\phi_{\mathrm{cut}}$ remains the same as in the
Einstein case~\cite{Hartle:2008ng}. Therefore, there will be a long
enough inflationary stage, and Einstein gravity will be recovered
when the inflation ends at $|\phi|<\phi_{\mathrm{cut}}$. For
example, a simple function like
$m_g^2=m_0^2\exp[-(\phi_{\mathrm{cut}}/\phi)^2]$ seems to satisfy
the requirement. Of course, however, we need a more thorough
analysis before we may conclude that such a model can actually lead
to the scenario described above (some discussions on its
dynamics have been done in some references, e.g.~\cite{LSS:2013}).

\noindent
(2) \textit{Running mass parameter}: Quantum effects may change $m_g^2$
through energy scales. If gravitational interactions are not asymptotically free,
then one may have a large graviton mass $m_g \sim M_{\mathrm{Pl}}$ at
the Planck energy scale, while it becomes small $m_g\sim H_0$
at the current energy scale, where $H_0$ is the current expansion rate of
the universe. If this scenario works, it may be also possible to explain the
accelerated expansion of the current universe simultaneously.

\section{Conclusion}

Studies of Hartle-Hawking no-boundary proposal for the wavefunction
of the universe in the context of dRGT massive gravity opens a
window to discuss inflationary scenario in quantum gravity theories.
Traditionally, the no-boundary wavefunction exponentially prefers
small number of $e$-foldings near the minimum of the inflaton
potential, and hence it does not seem to predict the universe we
observe today. However, we found that the contribution from the
massive gravity sector can drastically change this situation. We
showed that, for a fairly wide range of the parameters of the
theory, the no-boundary wavefunction can have a peak at a
sufficiently large value of the Hubble parameter so that one obtains
a sufficient number of $e$-folds of inflation.

To make this model to work, however, we need to find a way to trigger
the mass parameter in the very early universe while it should remain to
be extremely small in the current universe.
We speculated a couple of mechanisms for this purpose.
It is a future issue to see if these mechanisms can be actually
implemented in massive gravity.
In addition, we remain a future issue to consider the implication of potential problems of the dRGT model \cite{Deser:2012qx}
and applications for possible generalized massive gravity models that may not suffer from such problems, e.g., \cite{Lin:2013sja}.

\begin{acknowledgments}
This work was supported by the JSPS Grant-in-Aid for Scientific
Research (A) No.~21244033.
\end{acknowledgments}

\end{document}